\documentclass{article}

\usepackage{amsmath}
\usepackage{amssymb}
\usepackage[dvipsnames]{xcolor}
\usepackage{rotating}
\usepackage{natbib}
\usepackage{hyperref}
\usepackage{graphicx}

\usepackage[affil-it]{authblk}

\begin{document}
\title{International Trade Network: Country centrality and COVID-19 pandemic}

\author[1]{Roberto Antonietti\thanks{roberto.antonietti@unipd.it}}
\author[2]{Paolo Falbo\thanks{paolo.falbo@unibs.it}}
\author[1]{Fulvio Fontini\thanks{fulvio.fontini@unipd.it}}
\author[2]{Rosanna Grassi \thanks{rosanna.grassi@unimib.it}}
\author[2]{Giorgio Rizzini \thanks{giorgio.rizzini@unimib.it}}

\affil[1]{Department of Economics and Management "Marco Fanno", University of Padova, Padova, Italy}
\affil[2]{Department of Economics and Management, University of Brescia, Brescia, Italy}
\affil[3]{Department of Statistics and Quantitative Methods, University of Milano - Bicocca, Milan, Italy}

\date{}

\maketitle

\textbf{Keywords:} World Trade Network; Centrality Measures; Community Detection; COVID-19

{\textbf{JEL classification}}: F11;  F14; F62 \\
\smallskip
{\textbf{MSC 2010 classification}}: 91B60; 05C82; 62P20

\begin{abstract}
International trade is based on a set of complex relationships between different countries that can be modelled as an extremely dense network of interconnected agents. On the one hand, this network might favour the economic growth of countries, but on the other, it can also favour the diffusion of diseases, like the COVID-19. In this paper, we study whether, and to what extent, the topology of the trade network can explain the rate of COVID-19 diffusion and mortality across countries. We compute the countries' centrality measures and we apply the community detection methodology based on communicability distance. Then, we use these measures as focal regressors in a negative binomial regression framework. In doing so, we also compare the effect of different measures of centrality.\ Our results show that the number of infections and fatalities are larger in countries with a higher centrality in the global trade network.
\end{abstract} 

\section{Introduction}

\label{intro}

%(*** Logic steps: 1brief history ***)

At the beginning of 2020, the COVID-19 disease rapidly diffused from the
local Chinese region of Hubei, becoming soon a global health emergency.
%The spreading of the COVID-19 disease has severely affected all the economies in the world. 
Since it originated in a highly populated region,
strategic for several industrial sectors, the effects of lockdown
restrictions led to a freezing of business investments and a reduction in
Chinese household consumption, which had a significant impact on Chinese trades. Rapidly, the spreading of the COVID-19 disease has severely affected all the economies in the world.

Understanding the factors that triggered %favoured the first wave of 
the COVID-19 outbreak is still an object of debate. 
%(*** direct and indirect contagion factors ****)
%\color{paolo} 
The spread of a pandemic is a complex matter and %is subject to
can be affected by several factors. Much of the complexity is that many of these factors interact. 
In the case of COVID-19, the aerosol droplets spreading with
normal breathing and talking when two or more individuals meet physically,
has been identified as the main contagion mechanism. %The average level of temperature, humidity or the presence of particulate matter in the air have been found to condition the average time the virus survives. 
Keeping 1.5 or more meters of distance between individuals reduces the risk of contagion.
However, despite their necessity, physical explanations seem to provide only
a part of the reasons explaining the ongoing COVID-19 pandemic. %Surprisingly enough, they even seem to provide quite a secondary explanatory power. 
The reasons and the opportunities to bring people to travel around the world, to
spend time meeting other people seem more crucial to this purpose. Different
business, social and (or) family reasons discriminate much more effectively
the chances for people to enter physically in contact with each other, at
both global and local scales. Economic and social factors seem to provide
much more explanatory power to understand the COVID-19 pandemic. %

In this paper we stress the role of
international trade, and specifically of the countries' central position in the global trade scenario. From a topological point of view, the commercial trades between countries are characterized by an intricate weave of
relations, and the complex network theory offers an effective representation
of this situation. Both connections between countries and bilateral trade
flows can be modelled as a dense network of interconnected agents. Assessing the role of these strong interconnections on
the dynamics of the COVID-19 diffusion, and mortality, during the first
wave (between March and April 2020) is therefore an interesting research
question.

However, a major difficulty arises in the search of such
interconnections. More precisely, since the trade network is naturally dense and almost complete, the study of the classical global network
indicators applied to the whole network are not informative enough. This suggests an accurate choice of more effective network tools.

In this view, a first aim of the paper is %to identify more precise network centrality measures of countries and 
to assess the countries centrality as well as to identify 
a more satisfactory representation of the international trade landscape. Focusing on years 2019 and 2020, such analyses should allow us to detect and describe
more precisely if any changes in the international trade network have
eventually occurred as a consequence of the COVID-19. In doing so, we focus on those measures that are meaningful in capturing possible modifications.
A second aim is to assess if such centrality measures also have any explanatory power with
respect to the huge differences in the rate of infection, and mortality that
have been observed worldwide, once controlled for a series of other
confounding factors.

Some works %already appeared 
in the literature suggest that the level of
mobility of people (both at international and local levels) has played a
role in the COVID-19 pandemic. As pointed out in \cite{Antonietti2021} and \cite{Fernandez2020}, an intensive international mobility of people (for business or tourism reasons) can explain the
different infection and mortality rates across countries. More precisely,
countries with higher level of inward international mobility have higher
probabilities to anticipate the time of the first contagions and to have
higher number of infected people freely circulating during the pre-symptoms period. When on March 12th, 2020, the World Health Organization (WHO) announced the COVID-19 pandemic outbreak, both the contagion and mortality rates were already much different from country to country, with China and some EU countries already severely hit. \cite{Russo2020} %(*** PER FAVORE SISTEMARE
%IL FILE .BIB CI HO ANCHE PROVATO, MA io faccio casino, appoggio qui i
%riferimenti **** Russo L, Anastassopoulou C, Tsakris A, Bifulco GN, Campana
%EF, Toraldo G, et al. (2020) Tracing day-zero and forecasting the COVID-19
%outbreak in Lombardy, Italy: A compartmental modelling and numerical
%optimization approach. PLoS ONE 15(10): e0240649.
%https://doi.org/10.1371/journal.pone.0240649 ***)
point to January 18th as day-zero of the COVID-19 outbreak in Lombardy (Italy), which has been one of the mostly hit regions worldwide. \cite{parodi2020} %(*** APPOGGIO
%QUI Parodi E., Aloisi S. (2020). Italian scientists investigate possible
%earlier emergence of coronavirus, Reuters,
%https://www.reuters.com/article/us-health-coronavirus-italy-timing-idUSKBN21D2IG ***) 
suspect that the abnormal number of cases of bilateral pneumonia occurred in Lombardy already in December 2019 could be attributed to COVID-19. %
A factor that increases the probability of early contagion in
a region, or a country, is certainly the movement of the citizens outside
and inside its borders. In their cross-sectional analysis based on the
Spanish regions, \cite{paez2020} 
%Paez et al. (2020) (*** APPOGGIO qui. Paez A., Lopez F.A.,
%Menezes T., Cavalcanti R., Da Rocha Pinta M.G. (2020). A Spatio-Temporal
%Analysis of the Environmental Correlates of COVID-19 Incidence in Spain,
%Geographical Analysis, 0, 1-25) 
observe that local public mass
transportation system, more than international airport facilities, appears a
factor linked to higher severity of contagion rates. International and local transports seem to act differently. The former increases the chances of early contagion events, while the latter plays a second-order contagion enhancer.\\
International trade data can be used as a comprehensive indicator accounting for population density, economic dynamism, and human mobility. In this regards, \cite{bontempi2021bis} investigate the relations between the  total import and export of 107 provinces of Italy and the COVID-19 transmission dynamics. Extending the previous work, \cite{Bontempi2021} focus the regional data of France, Italy and Spain and confirm the relevance of trade in the analysis of COVID-19 pandemic, finding strong positive correlation between the international trade volume of each region and the percentage of patients recovered in the intensive care units. \\ 
%Such results further motivate the analysis of the link between COVID-19 pandemic and the trade networks among countries. 
From a network perspective, the impact of topology and metric properties on the stability and resilience of an economic or financial system has been widely studied in the literature, see e.g. \cite{Kali2007, Piccardi2018}. \\
On the one hand, community detection is an useful tool to see how an external shock modifies the topological structure of complex systems (\cite{Fortunato2016}).
%In the literature, two main categories appear: the algorithm-based and the optimization-based methods. In the first category are included methods based on hierarchical clustering or edge removal, as in \cite{Newman2004a}. The optimization-based method consists on determine the optimal network partition (over all possible ones) such that an optimization of specific criteria function is solved. We refer to \cite{Fortunato2010} and \cite {Fortunato2016} for an exhaustive review about community detection methods and algorithms.\newline
On the other hand, a suitable metric can highlight the role of non-local interactions between nodes. In this regard,  \cite{Estrada2008, Estrada2009} introduce the concept of communicability, presenting  
%The use of the communicability reveals that a given node is influenced not only by the exclusively interactions of its immediate neighbors but also by the interaction of the more distance ones.\newline Moreover, Estrada presents a new concept of communicability 
a metric between nodes that takes into consideration long-range interactions between them.\\ 
An area in which these concepts allow us to gain a deep insight into the hidden structures of the network is properly the World Trade Network (WTN), see \cite{Bartesaghi2020}. \\
The topology of the WTN has been extensively analyzed over time. The behavior of international trade flows, the impact of globalization on the international exchanges, the presence of a core-periphery structure or the evolution of the community centres of trade, are just some of the issues addressed by the recent developments (see  \cite{Serrano2007, Tzekina2008, Fagiolo2010, DeBenedictis2011, Blochl2011, Grassi2021}).  
Recently, some works correlate the commercial trades with the COVID-19 diffusion from a network point of view (\cite{Antonietti2021bis, reissl2021,kiyota2021,fagiolo2020}.\\
Such results further motivate the analysis of the link between COVID-19 pandemic and the trade networks among countries. 
%Many works have dealt with the network from a multi layers perspective \cite{Snyder1979, Barigozzi2011} or aim to emphasize financial implications of the world trade or contagion processes on the network \cite{Wilhite2001, Reyes2008, Schiavo2010,Fagiolo2013, Fan2014, Varela2015, Giudici2016, DeBenedictis2016, Cepeda2019, Cerqueti2019}.
%(*** Contribution to the literature ***)

Our contribution to the literature is twofold. Firstly, we detect more precisely the presence of trade communities not only via their direct connections, as measured by the total volume of trade directly exchanged between two countries, but also via indirect connections. Indeed, we argue that it is crucial to consider deep interconnections between nodes, in order to capture the strategic commercial links, which can survive beyond a global shock. To this end, we apply the recent methodology proposed by \cite{Bartesaghi2020} focusing on the Estrada communicability distance (\cite{Estrada2009}). 
As a result, the analysis of communities performed in 2019 and 2020 shows that the trade network is a resilient structure, adapting itself to a global shock such as a pandemic. 
%Secondarily, we provide a first empirical evidence that COVID-19 has not generated significant changes in the international trade network. 
Secondly, we provide strong empirical
evidence that, on the contrary, the network centrality measures have
impacted to the early diffusion and mortality rates of COVID-19.
We show in particular that a higher country centrality in the WTN corresponds to a higher risk of infection and death. 
Also, the community clustering coefficient, that synthesizes both the community structure and countries centrality, leads to a higher number of deaths and infections with respect to the ones computed with the classical clustering coefficient. 

%To this end, we investigate the network structure and we run an econometric analysis based on two main data sources. First, we examine the global economic trade of countries in the first semester of 2019 and of 2020. (*** spostare questo nella sezione "data": Data are provided by the United Nations Statistical Division \cite{UN Comtrade}. ***)
%Such arguments highlight that besides being a source of wealth, international trade networks can be a source of risk for our modern global society. \\
The paper is organized as follows: in section 2 we describe the methodology
and the network indicators, as well as the econometric model used to perform the analysis. In section 3 we
describe the WTN and socio-economic data used and how we constructed the
world trade network. In section 4 we report and discuss the results of the
network analysis and the econometric model. Conclusions follow in section 5.

\section{Methodology and network indicators}
\label{Methodology}
In this section we describe the methodology that we  apply in the paper, in particular the community detection method, as well as the centrality measures used later for the econometric analysis.

Firstly, we briefly remind some preliminary  definitions. A network is formally represented by a graph $G=(V,E)$ where $V$ and $E$ are the sets of $n$ nodes and $m$ edges (or links), respectively. We consider simple graphs without self-loops.
Two nodes $i$ and $j$ are adjacent if there is an edge $(i,j)\in E$ connecting them. If $(i,j) \in E$ implies $(j,i) \in E$ then the network is undirected. A walk of length $k$ is a sequence of adjacent vertices $v_0,v_2,...,v_k$. When all vertices in a walk are distinct, the walk is a path. A $i-j$ geodesic is the shortest path between vertices $i$ and $j$. The length of the $i-j$ geodesic is the distance $d(i,j)$ between $i$ and $j$.
% 
%The shortest path, or geodesic, between $i$ and $j$ is a path with the minimum number of edges. The length of a geodesic is called geodesic distance or shortest path distance $d(i,j)=d_{ij}$. A graph $G$ is connected if, $\forall i,j \in V$, a $i-j$-path connecting them exists. \\
Adjacency relationships are represented by the adjacency matrix $\textbf{A}$, whose elements $a_{ij}$ are equal to 1 if $(i,j) \in E$ and zero otherwise.
We denote with $\lambda_1\geq \lambda_2\geq \dots \geq \lambda_n$ the eigenvalues of $\textbf{A}$ and $\rho= \max_i{(|\lambda|_i)}$ its spectral radius. By the Perron-Frobenius Theorem \cite{Horn2012} $\rho$ is an eigenvalue of $\textbf{A}$ and there exists a positive eigenvector $\mathbf{x}$ such that $\textbf{Ax}=\rho\textbf{x}$. Such an eigenvector $\textbf{x}$ is called principal eigenvector.\\ 
The $ij$-entry of the $k^{th}$ power of
the adjacency matrix $\textbf{A}^k$ provides the number of walks of length $k$ starting at $i$ and ending at $j$. We define the degree $d_i$ as the number of edges adjacent to $i$. 

%The degree $d_{i}$ of a node $i$ is the number of edges incident on it. 
%The diagonal matrix whose diagonal entries are $d_i$ is  $\textbf{D}$.

A graph $G$ is weighted when a positive real number
$w_{ij}>0$ is associated with the edge $(i,j)$. 
In this case, the adjacency matrix is a non-negative matrix $\textbf{W}$. 
We define the strength $s_i$ as the sum of the weights of the edges adjacent to $i$ and the diagonal matrix whose diagonal entries are $s_i$ is $\textbf{S}$. 
%When $w_{ij}=1$ if $(i,j) \in E$, then the
%graph is unweighted. Thus, the unweighted case can be viewed as a particular weighted one.\\

\subsection{Community detection based on communicability distance}
\label{Comm_method}

We describe here the community detection method using the Estrada communicability distance proposed in \cite{Bartesaghi2020}. %For further details, see \cite{Bartesaghi2020}.\\
We focus here on the main steps of the methodology. For a detailed description of the proposed methodology, we refer the reader to \cite{Bartesaghi2020}.\\ 
The main idea is to detect communities by optimizing a quality function that exploits the additional information contained in a metric structure based on the Estrada communicability. 
At first, we recall the definition of Estrada communicability (simply, communicability) between two nodes $i$ and $j$ (see \cite{Estrada2008}): 

\begin{equation}
\label{Ecomm}
G_{ij}=\sum_{k=0}^{+\infty}\frac{1}{k!}[\textbf{A}^k]_{ij}=\left[ e^{\textbf{A}} \right]_{ij}.
\end{equation}

As the $ij$-entry of the $k$-power of the adjacency matrix $\textbf{A}$ provides the number of walks of length $k$ starting at $i$ and ending at $j$, $G_{ij}$ accounts for all channels of communication between two nodes, giving more weight to the shortest routes connecting them. The elements $G_{ii} $, $i=1,...,n$ are known in the literature as subgraph centrality \cite{Estrada2005}. The communicability matrix is then the exponential of the matrix $\mathbf{A}$, simply denoted by $\textbf{G}$.

In the case of a weighted network, the weighted communicability function is defined as

\begin{equation}
\label{wecomm}
G_{ij}=\sum_{k=0}^{+\infty}\frac{1}{k!}[(\textbf{S}^{-{\frac{1}{2}}}\textbf{W}\textbf{S}^{-{\frac{1}{2}}})^k]_{i j}=\left[ e^{(\textbf{S}^{-{\frac{1}{2}}}\textbf{W}\textbf{S}^{-{\frac{1}{2}}})} \right]_{ij}
\end{equation}

%where $\textbf{S}$ is the diagonal matrix whose diagonal entries are the strengths of the nodes. 

Following \cite{Crofts2009}, the matrix $\textbf{W}$ in formula (\ref{wecomm}) has been normalized to avoid the excessive influence of links with higher weights in the network. 

Using the communicability, a meaningful distance metric $\xi_{ij}$ can be constructed, as defined in \cite{Estrada2012}:

\begin{equation}
\label{communicabilitydistance}
\xi_{i j}=G_{ii}-2G_{ij}+G_{jj}.
\end{equation}

By definition of communicability, $G_{ij}$ measures the amount of information transmitted from the node $i$ to $j$. On the other hand, $G_{ii}$ measures the importance of a node according to its participation in all closed walks to which it belongs. Hence, in terms of information diffusion, $G_{ii}$ is the amount of information that, after flowing along closed walks, returns to the node $i$. 

Thus, the quantity $\xi_{ij}$ accounts for the difference in the amount of information that returns to the nodes $i$ and $j$ and the amount of information exchanged between them. 
The greater is $G_{ij}$, the larger the information exchanged and the nearer are the nodes; the greater are $G_{ii}$ or $G_{jj}$, the larger the information that comes back to the nodes and the farther are the nodes. 
Since $\xi_{i j}$ is a metric, then $G_{ii}+G_{jj}\geq 2G_{ij}$, i.e., no matter what the structure of the network is, the amount of information absorbed by a pair of nodes is always larger or equal than the amount of information transmitted between them. \\
The metric is meaningful if we apply it to international trade network. Indeed, network flows along links measure how well two countries communicate in terms of commercial exchanges. For instance, the link between two nodes may be identified with the total trade or money flow between two countries. 

We assume that two nodes are considered members of the same community if their mutual distance $\xi_{ij}$ is lower than a threshold $\xi_0 \in [\xi_{min},\xi_{max}]$. 
In particular, we construct a new community graph with adjacency matrix $\textbf{M}=[m_{ij}]$ given by:
\begin{equation}
m_{ij}=
\left\{ 
\begin{array}{ll}
1 & \ {\rm if}\ \xi_{ij}\leq \xi_0 \\ 
0 & \ {\rm otherwise} \\ 
\end{array}
\right.
\label{matrixM}
\end{equation}

In this way, clustered groups of nodes that ``strongly communicate'' emerge, varying the threshold $\xi_0$.

As well explained in \cite{Bartesaghi2020}, $\xi_0$ is not arbitrarily chosen, but it is obtained by solving the following optimization problem 
$$
\xi_0 \in \arg \max Q.
$$
The objective function $Q$ is 

\begin{equation}
\label{Qu}
Q = \sum_{i,j} \gamma_{ij}x_{ij},    
\end{equation}

where $x_{ij}$ is a binary variable equal to $1$ if nodes $i$ and $j$ belong to the same community and $0$ otherwise. 
$\gamma_{ij}$ is a function measuring the cohesion between nodes $i$ and $j$. Originally proposed in \cite{chang2016}, it is defined in \cite{Bartesaghi2020} as:

\begin{equation}
\gamma_{ij} = (\bar{\xi}_j - \bar{\xi}) - (\xi_{ij}- \bar{\xi}_i),    
\end{equation}

where $\bar{\xi}_{j}$ is the average distance between node $j$ and nodes other than $j$ and $\bar{\xi}$ is the average distance over the whole network.
 
Since two nodes are cohesive (incohesive, respectively) if $\gamma_{ij}\geq 0$ $(\gamma_{ij}\leq0)$, in terms of distance they are cohesive if they are close to each other and, on average, they are both far away from the other nodes.\\
In this perspective, $\gamma_{ij}$ can be seen as the ``gain'' if positive or the ``cost'' if negative, of grouping two nodes $i$ and $j$ in the same community. 
The applied methodology will allow to discover communities in the world trade network based on all the possible channels of interactions and exchange between countries.

\subsection{Centrality measures}
In this section we introduce the centrality measures that will be used in performing the econometric analysis. 

The centrality of a node indicates its importance  in the network \cite{Sabidussi1966}. There are several definitions of vertex centrality in a network, depending on the application. Among them, some measures seem to better capture the characteristics of the studied network.\\ 
The degree centrality, formally expressed by the degree $d_i$, quantifies the ability of a node $i$ to communicate directly with others and it is 
\begin{equation}
\label{degree_meas}
d_i = \sum_{j=1}^n a_{ij}.
\end{equation}
For weighted networks with weights matrix $\mathbf{W}$, a similar measure is the strength centrality $s_i$,
\begin{equation}
\label{strenght_meas}
s_i = \sum_{j=1}^n w_{ij}.
\end{equation}
An extension of the degree centrality is the eigenvector centrality (see \cite{Bonacich1972}). 
Formally, it is represented by the $i$-th component of the principal eigenvector $\mathbf{x}$ of the adjacency matrix:
\begin{equation}
\label{eigenvalues_meas}
x_i = \frac{1}{\rho} \sum_{j=1}^n a_{ij} \, x_j
\end{equation}
The eigenvector centrality $x_i$ quantifies the connection of a vertex with its neighbours that are themselves central. The extension to weighted case is immediate, as the weighted adjacency matrix $\mathbf{W}$ preserves all characteristics of $\mathbf{A}$.\\
Betweenness centrality $b_i$, measures the influence
that a vertex $i$ has in the spread of information within the network:
\begin{equation}
\label{betweenes_meas}
    b_i = \sum_{h,k} \frac{g_{hk}(i)}{g_{hk}}, \hspace{2mm} h,k \ne i
\end{equation}

where $g_{hk}$ is the number of $h-k$ geodesic from $h$ to $k$ and $g_{hk}(i)$ is the number of $h-k$ geodesic passing through node $i$. In the definition of betweenness we always suppose that the pairs $(i,j)$ appear only once in the sum.\\

We also consider the local clustering coefficient, which measures the tendency to which nodes in a network tend to cluster together.
Since the world trade network will be represented as an indirect and weighted network (see Section \ref{Dataset_network}), we focus on the local weighted coefficient proposed by \cite{Onnela_2005}:
\begin{equation}
\label{onnela_meas}
C_i(\mathbf{\tilde{W}}) = \frac{\sum_{j} \sum_{j\ne k} \tilde{w}_{ij}^{1/3} \tilde{w}_{jk}^{1/3} \tilde{w}_{ki}^{1/3}}{d_i(d_i -1)}
\end{equation}
where $\mathbf{\tilde{W}}$ is the weighted adjacency matrix obtained by normalizing the entries $w_{ij}$ of $\mathbf{W}$ as $\tilde{w}_{ij} = \frac{w_{ij}}{\max(w_{ij})}$ $\forall i,j$. 
Notice that $C_i(\mathbf{\tilde{W}})=C_i$ represents the geometric mean of the links weights incident to the node $i$, divided by the number of potential triangles $d_i$ centred on it. The main idea is to replace the total number of the triangles in which a node $i$
belongs, with the ``intensity'' of the triangle, defined here as the geometric mean of its weights. \\

\subsection{Econometric model}
\label{econometric_model}
In what follows, we want to assess the role of the WTN on the evolution of the pandemic in the five weeks between March 11th and April 21st 2020. At the same time, we want to control for additional socio-economic factors that can have an impact on the diffusion of the pandemic. To avoid the possibility that, in turn, these factors might be affected by COVID-19 diffusion, we include them as referred to year 2019. \\
The baseline model that we adopt to test for the role that Trade Network Centrality has played in explaining the number of infections (INF) and deaths (DEATH) in the first wave of the COVID-19 outbreak (i.e., between March 11st, 2020 and April 21st, 2020) is the following: 
\begin{equation}
\label{econometric1} 
Y_{it}=\beta_0 + \beta_1 TNC_{i}+\mathbf{Z^{'}}_{i}\beta_{Z}+\gamma_t + \epsilon_{it}
\end{equation}
where $Y_{it}$ \color{black} is either the number of COVID-19 infections (INF) or the number of deaths (DEATH) in country $i$ and week $t$. The variable $TNC_{i}$ \color{black} represents a given trade network centrality measure (respectively: degree in Equation \eqref{degree_meas}, betweenness in Equation \eqref{betweenes_meas}, local clustering coefficient in Equation \eqref{onnela_meas}, weighted eigenvector in Equation \eqref{eigenvalues_meas}, and strength in Equation \eqref{strenght_meas}) measured in 2019; $\mathbf{Z}$ is a vector of additional regressors that can explain the number of infections and fatalities due to COVID-19, namely GDP per capita (GDPPC, at constant 2010 US\$), total resident population (POP), the share of elderly population (POP65+), the number of hospital beds per 1,000 inhabitants (HBEDS), and the average temperature in February and March (in degree Celsius), all measured in 2019. The term $\gamma_t$ is a vector including a set of five week-specific dummies that capture the trend in the dynamics of COVID-19 infections and fatalities for all our countries, while $\epsilon_{it}$ is the stochastic error component with zero mean and finite variance $\sigma^2_{\epsilon}$. Since the residuals of Equation \eqref{econometric1} are likely to be correlated within countries, we cluster the standard errors at the country level. 
Since $Y_{it}$ is a count variable, and our regressors are time-invariant because they are all measured in 2019, we estimate Equation \eqref{econometric1} using a pooled negative binomial regression model. 
As common for count-data models, we test for the overdispersion of our data, that is, for the fact that the conditional mean can be lower than the conditional variance, typically due to the presence of unobserved factors than can affect the number of COVID-19 infections or deaths. In such a case, the main assumption for the use of the Poisson model is violated, and the negative binomial model is more suited to estimate Equation \eqref{econometric1}. \\
We also check for the presence of potential multicollinearity by re-estimating Equation \eqref{econometric1} through a linear regression model and using a Variance Inflation Factor (VIF) statistic \footnote{The VIF is the ratio of variance in a  model that uses multiple independent variables and the variance of a model that uses only one independent variable. This statistic is used to test for the severity of multicollinearity in linear regressions. For see in detail how VIF statistic works, we refer for instance to \cite{james2013}.}.
%The test works in three steps. First, we run a linear regression. 
%where each independent variable $z$ of the original model is regressed against all the remaining ones. Second, we compute the VIF score from each regression as follows VIF$_{z}=\frac{1}{(1-R^2_{j})}$. Third, we compare each VIF$_{z}$ with a cutoff value, which, in the strictest interpretation, is set to 5. Values of the VIF above $5$ signal a possible problem of multicollinearity in the original linear regression model.} 
Multicollinearity can be considered an issue if the VIF statistic takes a value higher than the commonly accepted threshold of 5. To check which of the proposed trade network centrality measures provides the highest explanatory power in predicting $Y_{it}$ we use the Akaike Information Criterion (AIC) and Bayesian Information Criterion (BIC). \\
Finally, to compare the magnitude of the estimated coefficients, we standardize all the regressors by subtracting their mean and dividing by their standard deviation. For each variable, we report the incidence rate ratio (IRR) which measures the impact of a unit increase of the regressor on the risk of contagion (mortality) from COVID-19, computed as the ratio between the number of infected (deceased) individuals and the number of non-infected (surviving) individuals. In this respect, the IRR of a regressor is easier to interpret than the corresponding estimated coefficient, since this latter is the impact of a unit increase in the regressor itself on the log of the expected number of infections or deaths. 

\section{Data, samples and variables}
%In this section we formalize the international trade using network theory.
\label{Dataset_network}

\subsection{Dataset description}
The empirical analysis is based on two datasets. The first is used to construct the network and consists of a sample of monthly trade values during the first semester of 2019 and of 2020 for $55$ countries listed in Tables \ref{tab:countries1} and \ref{tab:countries2} and  in the Appendix 1. %The observation times are the first semester of 2019 and of 2020.\ 
Data are provided by the \cite{UNComtrade}, that is the largest depository of international trade data. It contains over 40 billion data records since 1962 and is available publicly on the internet.\\ 
The second is used to analyze the relationship that such a trade network has with COVID-19 diffusion.
Data on COVID-19 diffusion come from the \cite{ECDPC} (ECDC), an EU agency for the protection of European citizens against infectious diseases and pandemics. The data on the distribution of COVID-19 worldwide are updated daily by the ECDC’s Epidemic Intelligence team, based on reports provided by national health authorities. Since we are interested in the first wave of the pandemic, we retrieve cross-country daily data on the number of COVID-19 infections and deceases, that we pool into five weeks from March 11st, 2020 to April 21st, 2020. \\
To control for other factors that can potentially affect the diffusion patterns of COVID-19, we also consider the following country-level information provided by \cite{WB}: the real GDP per capita (GDPPC, in 2010 USD) used as a proxy for the average standard of living in a country; the total resident population (POP), taken as a proxy for a country's size; the share of population aged 65 or more (POP 65+); the number of hospital beds per capita available in public, private, general, and specialized hospitals and rehabilitation centers (HBEDS), that we include to capture the average quality of the health system in each country; the average temperature in February and March (TEMP), in degree Celsius, C. \\

Tables \ref{Summary statistics} and \ref{Correlation matrix} show the main summary statistics and the pairwise correlations among regressors.

\begin{table}[h]
\centering
\begin{tabular}{l c c c c} 
 \hline
\textbf{Variable} & \textbf{Mean} &	\textbf{Std. dev.} &	\textbf{Min} &	\textbf{Max} \\ 
 \hline
 \emph{Network centrality} & & & &  \\
 \hline
 Degree	&50.62	&5.223	&27	&54\\
 Betweenness &	0.0011 &	0.0008 &	0&	0.0019\\
Local clustering  &	0.0022&	0.0026&	0.0001&	0.0122\\
Weighted Eigenvector	&0.109&	0.198&	0.0002	&1\\
Strength (:$10^9$)	&53.71&	96.00&	0.130&	485.9\\
\hline
\emph{Additional regressors} &&&&\\
\hline
GDPPC&	27085.33&	26413.5&	809.36&	111062.3\\
POP (mln)&	53.754&	187.62&	0.3613&	1366.4\\
POP65+&	0.147&	0.063&	0.026&	0.280\\
HBEDS&	4.031&	2.375&	0.600&	13.40\\
TEMP ($C$)&	6.181&	10.89&	-20.99&	26.80\\
\hline
\end{tabular}
\caption{Summary statistics}
\label{Summary statistics}
\end{table}

\begin{center}
\begin{sidewaystable}
%\begin{table}
\centering
\begin{tabular}{l c c c c c c c c c c} 
 \hline
 & Degree&	Betweenness&	Clust. coeff.&	W Eigen.&	Strength&	GDPPC&  POP&	POP65+&	HBEDS&	TEMP\\
\hline 
1. Degree &	1& & & & & & & & & \\							2. Betweenness	&0.76***&	1& & & & & & & & \\
3. Clust. coeff. &0.45***&0.62***&1& & & & & & & \\
4. W Eigen.&0.33***&0.50***&0.87***&1& & & & & & \\
5. Strength	& 0.34***&0.50***&	0.96***&	0.95***&	1& & & & &  \\
6. GDPPC &0.48***&0.59***&0.46***&0.42***&0.40***&1& & & & \\	7. POP &0.15***&0.23***&0.23***&0.23***&0.21***&-0.11&1& & &  \\
8. POP65+ &0.61***	& 0.51***	& 0.44***	& 0.32***	& 0.35***	& 0.51***	& -0.19***	& 1	& & \\
9. HBEDS &	0.36***&	0.16***&	0.25***&	0.15***&	0.20***&	0.20***&	0.20***&	0.66***&	1 & \\
10. TEMP &	-0.44***	&-0.17***	&-0.22*** &	-0.29***&	-0.24***&	-0.36***&	0.19***&	-0.59***&	-0.54***&	1 \\
\hline
\multicolumn{11}{l}{\emph{*** Significant at 1\% level.}}
\end{tabular}
\caption{Correlation matrix among regressors}
\label{Correlation matrix}
%\end{table}
\end{sidewaystable}
\end{center}

%SPOSTARE PIU' AVANTI
%As discussed in the Introduction, part of this work consists in the analysis of the communities in the World trade network. %at different times. 
%\footnote{The analysis has been performed for a small sample of countries since for the first semester of $2020$ at the download time data are missing for many of the biggest countries in the international trade, such as France, Russia, China.}\\
%because of lack of data (at the download time) about the first semester of $2020$. \\

\subsection{Network construction}
Trades between countries are represented as a weighted network, where each country is a node and connections, i.e. links between nodes, is measured by the amount of traded volume (expressed in US-dollars).\\
At first, we compute separately the aggregate trade values of import and export between each pair of countries. 
%Then, we put a link between two countries if both import and export exist. 
Then, we consider a pair of countries $(i,j)$ such that both import and export exist. Specifically, focusing on trade flows from $i$ to $j$, let $w^{imp}_{ij}$ and $w^{exp}_{ij}$ be the aggregate import trade value and the aggregate export value, respectively, from $i$ to $j$. We then put a weighted link from $i$ to $j$ representing the average value between import and export, defined as follows:

$$
\bar{w}_{ij}= 
\begin{cases}
\frac{w^{imp}_{ij}+w^{exp}_{ij}}{2} & \text{if $w^{imp}_{ij}>0$} \text{and $ w^{exp}_{ij}>0$} \\
0 & \text{otherwise}
\end{cases} 
$$
Notice that, due to incompleteness of data\footnote{Since we refer to the years 2019 and 2020 not all the countries have completely communicated the data to UN Comtrade}, in general $\bar{w}_{ij} \neq \bar{w}_{ji}$. The resulting network is then weighted and oriented, with eventually bilateral links between two nodes. 
%Reasonably, we expect that the aggregate import trade from node $i$ to node $j$ equates the aggregate export trade from node $j$ to node $i$ so that an indirect network can be constructed. In our case, incompleteness of data information and data missing make the network oriented. \\

Since the approach of \cite{Bartesaghi2020}, based on the communicability distance, has been developed on indirect networks, we investigate if it is possible %\color{fulvio} non capisco cosa intendete dire con we keep if it possible to neglect. intendete whenever possible, we neglect ...?  \color{black} 
to neglect the direction of the links. To this end, we compute the Spearman correlation between in and out strength distribution for each year of the sample. The resulting correlations is $0.99$ for both years.  
We then can substitute the bilateral arcs between nodes $i$ and $j$ with a single 
non-oriented link having weight given by 
the maximum value between $\bar{w}_{ij}$ and $\bar{w}_{ji}$, i.e. 
$$
w_{ij} = \max(\bar{w}_{ij},\bar{w}_{ji}).
$$ 

This choice is based on an information quality reason: we expected that the higher is the value traded, the bigger is the information that can contain. 

In Table \ref{tab:table_network} we report the global network indicators referred to the WTN for years $2019$ and $2020$. %Preliminary analysis shown in Table \ref{tab:table_network} on the WTN for $2019$ and $2020$ years
%Density is defined as the ratio between the average degree and the maximum degree ($n-1$). INSERIRE TRANSITIVITY

\begin{figure}[h]
\includegraphics[scale=0.50]{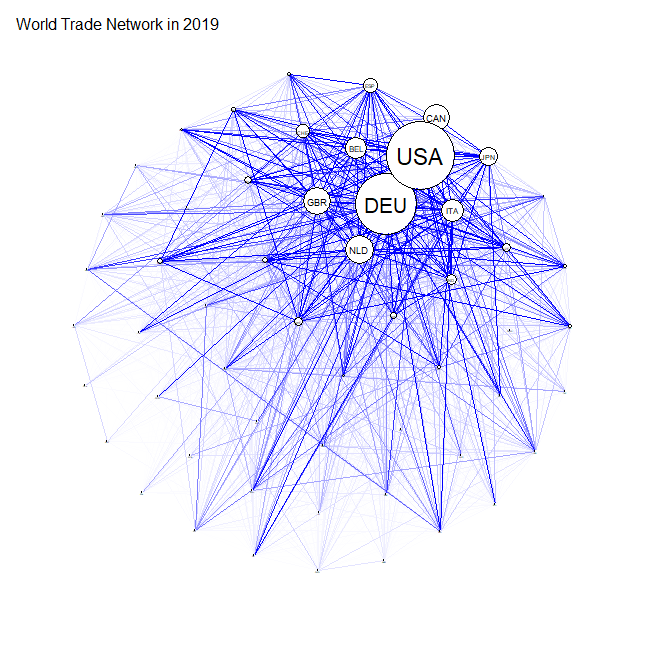}
\caption{World Trade Network representation in 2019. The size of the nodes is proportional of its strength.}
\label{m_network_2019}
\end{figure}

\begin{figure}[h]
\includegraphics[scale=0.5]{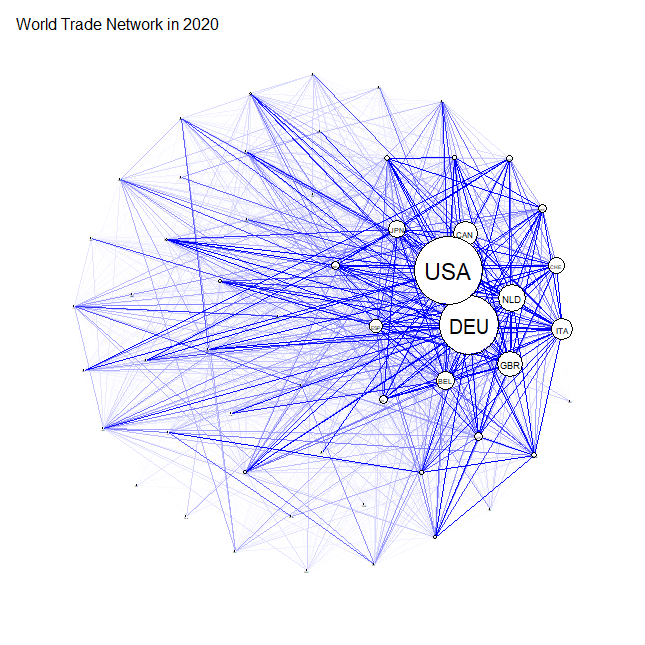}
\caption{World Trade Network representation in 2020. The size of the nodes is proportional of its strength.}
\label{m_network_2020}
\end{figure}

As expected, the network shows an extremely connected, dense and almost complete structure. This is typical of this kind of network, being the economic trades pervasive in all the world. 
This is certainly confirmed by the average degree ($51$ and $52$) and density\footnote{Density measures how many links between nodes exist compared to how many links between nodes are possible.} ($0.937$ and $0.948$ for $2019$ and $2020$, respectively). A high value of the transitivity ($0,953$ and $0.959$ for $2019$ and $2020$, respectively)\footnote{The transitivity coefficient is the ratio between the number of actual triangles and the number of the potential ones, %introduced by Newman in 2001
see \cite{newman2001}. It expresses the network cliquishness, as it can be seen as the probability that the adjacent nodes of a reference nodes are themselves connected} denotes a strong interconnection among countries. 
The WTN network is depicted in Figures \ref{m_network_2019} and \ref{m_network_2020}.\\

%In this circumstance, we will focus on the local network indicator since the classical global ones are not informative. 

\begin{table}[h]
    \centering
    \begin{tabular}{lcc}
    \hline
         Feature & $2019$ & $2020$ \\
         \hline \hline
         Number of nodes& $55$& $55$\\ 
         Number of links& $1392$& $1409$\\
         Average degree & $51$ & $52$\\
         Density & $0.937$ & $0.948$\\
         Transitivity& $0.953$ & $0.959$\\
         %Assortativity& $-0.06$ & $-0.021$\\
         \hline
    \end{tabular}
    \caption{Global network indicators of World Trade Network for years $2019$ and $2020$.}
    \label{tab:table_network}
\end{table}

\section{Results}

\subsection{Evolution of WTN during COVID-19}

We apply the methodology described in Section \ref{Comm_method} by using the communicability
distance. As already stressed in the previous Sections, WTN is characterized by an almost complete structure, where direct connections between countries are dominant. By this approach we have a tool to quantify the depth of the level of communication between countries.
Indeed, as pointed out in \cite{Bartesaghi2020}, in the WTN two countries are directly connected by a link in terms of products they directly exchange. However, a higher order exchange may occur between them, for instance when they are involved in a chain of production. The communicability distance allows us to to take into account trades between countries that take place indirectly.

At first, we compute the communicability distance between countries for both networks, referred to years 2019 and 2020, applying Formula (\ref{communicabilitydistance}). 
%Computing the Estrada communicability in the WTN 2019, we obtain that 
In 2019 the nearest countries are Canada and United States ($\xi_{\min}= 1.1166$) and they still remain in 2020 ($\xi_{\min}= 1.1316$). In 2019 the farthest countries are Kyrgyzstan and United States of America ($\xi_{\max}=1.4969$), whereas in 2020 the farthest are United States and Iceland ($\xi_{\max}=1.5077$). These results underline the central role of the United States in the WTN also at the beginning of a pandemic.

\begin{figure}[h]
\includegraphics[scale=0.4]{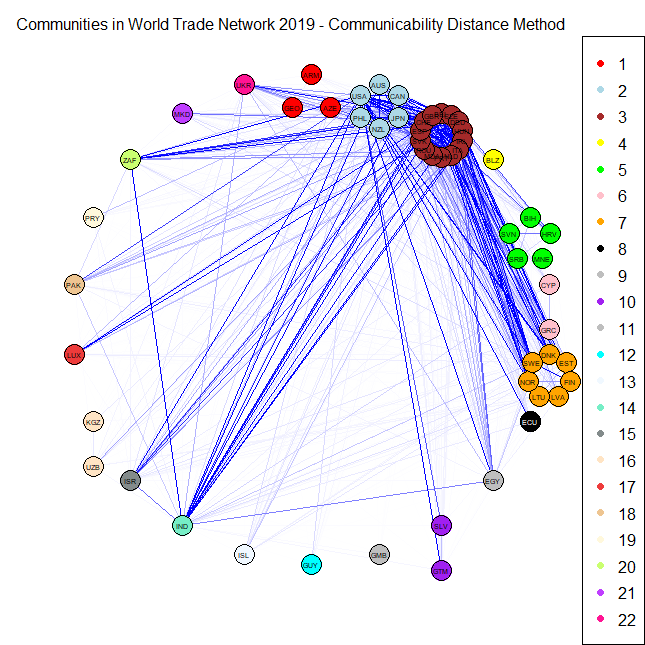}
\caption{Community graph obtained with communicability distance method in $2019$.}
\label{m_figure_BCG_2019}
\end{figure}

\begin{figure}[h]
\includegraphics[scale=0.05]{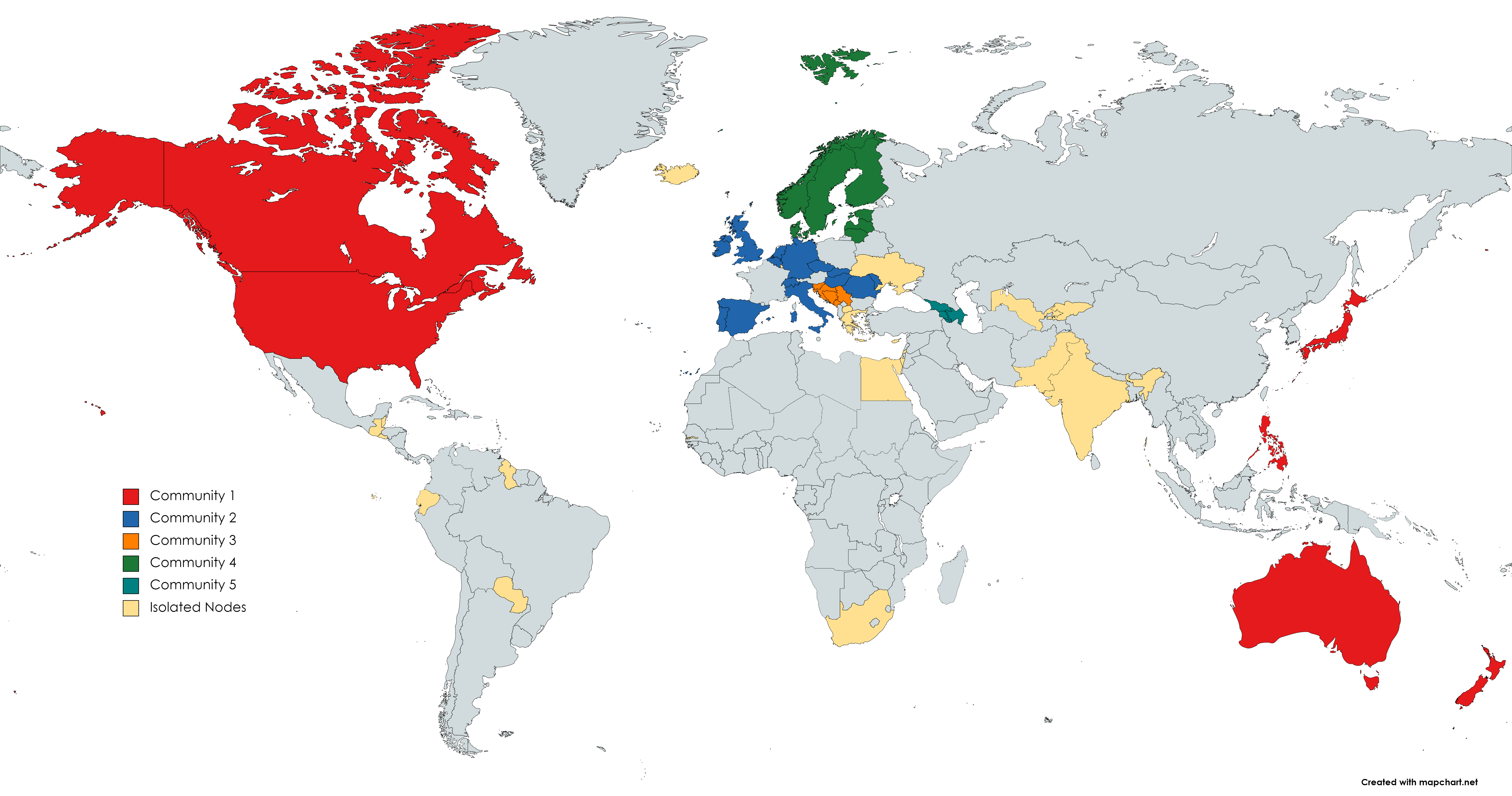}
\caption{World-map with the optimal community structure in $2019$.}
\label{BCG_2019}
\end{figure}

\begin{figure}[h]
\includegraphics[scale=0.4]{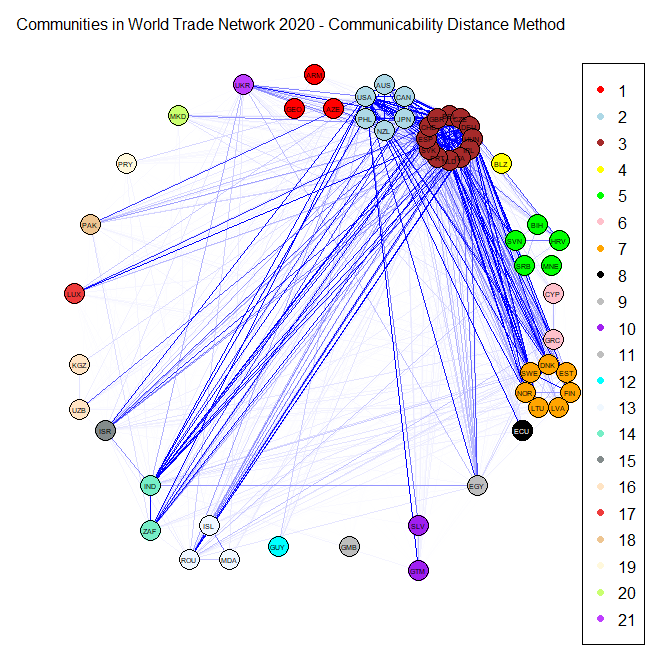}
\caption{Community graph obtained with communicability distance method in $2020$.}
\label{m_figure_BCG_2020}
\end{figure}

\begin{figure}[h]
\includegraphics[scale=0.05]{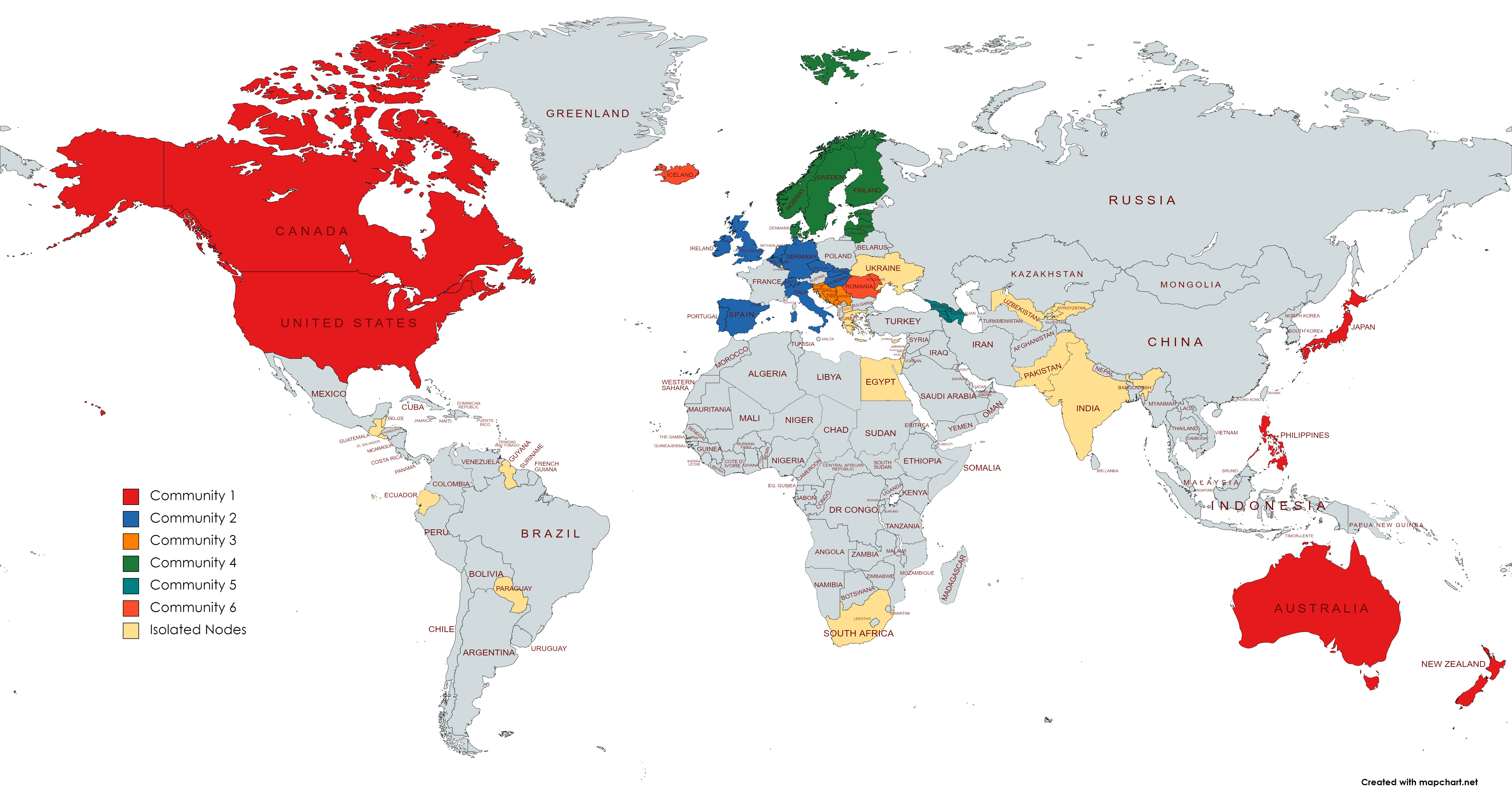}
\caption{World-map with the optimal community structure in $2020$.}
\label{BCG_2020}
\end{figure}

We then apply the community detection method based on the communicability distance. The optimal threshold $\xi_0^*$ maximizing the quality function $Q$ defined in Equation \eqref{Qu} is 1.3676 and 1.3723 for $2019$ and $2020$, respectively. The corresponding optimal value of the quality function $Q^*$ is 86.2301 and 87.0466 for $2019$ and $2020$.
Figures \ref{m_figure_BCG_2019} and \ref{m_figure_BCG_2020} report the community graphs whose adjacency matrices are computed according to Formula \eqref{matrixM}.   %of the community detection  i.e. matrix $\mathbf{M}$ of Section \ref{Methodology}, obtained applying the communicability distance method.
Communities obtained with the optimal threshold $\xi_0^*$ for both years are also shown in Figures \ref{BCG_2019} and \ref{BCG_2020}, through a world-map representation.\\

Grey countries are those not included in the sample while the yellow countries are the isolated nodes, which, in the following, will be denoted as the rest of the word. \\
%\color{giorgio} Chi ha messo questo commento? tenete presente che poi quando si stampa in b/n i colori non si vedono. forse meglio distinguerli con bold, italics, ecc. inoltre. In generale nel paper ci sono moltissime figure e tabelle. forse troppo. per esempio, non e' meglio lascire qui le figure, e mettere le tabelle corrispondenti in appendice? o vice versa?  \color{black} \\
Observe that in $2019$ the number of communities is $22$ with $18$ isolated nodes. whereas in $2020$, this  number is $21$ with $16$ isolated nodes. The slight reduction in the number of communities can be explained by the effect of a strengthening of the long-range alliances: on the one hand, countries reinforce the existing links and on the other hand, they make deals with geographically neighbors.\\
In correspondence of the optimal threshold $\xi_0^*$%\footnote{We notice that moving the threshold, one can obtain different community structure.}
, community detection on WTN shows a fragmented structure, where, among all, three strong communities emerge. %The rest of the section is devoted to study communities composition for both years. 
The first community is the European community, containing $12$ European countries the second one contains the United States, Canada, Japan, Australia and the third community sees the North European group. Member of those communities are listed in Table \ref{tab:CD2019BCG} for $2019$ and in Table \ref{tab:CD2020BCG} for $2020$.
\\

\begin{table}[h]
	\centering{}
	\begin{tabular}{|lll|}
		\hline \hline
		& \bf Size & \bf Members \tabularnewline
		{\color{red} \bf Community 1}  & \bf 6 & {\color{red} \small \bf AUS CAN JPN NZL PHL USA}\tabularnewline
		\hline
		{\color{blue} \bf Community 2} & \bf 14 & {\color{blue} \small \bf BEL CHE CZE DEU ESP GBR HUN} \tabularnewline
		{\color{blue} \bf } & & {\color{blue} \small \bf IRL ITA MDA NLD PRT ROU SVK }\tabularnewline
		\hline                             
		{\color{orange} \bf Community 3} & \bf 5 & {\color{orange} \small \bf BIH HRV MNE SRB SVN}  \tabularnewline
		\hline
		{\color{ForestGreen} \bf Community 4}& \bf 7 & {\color{ForestGreen} \small \bf DNK EST FIN LTU LVA NOR SWE}	 \tabularnewline
		\hline
		{\color{PineGreen} \bf Community 5}& \bf 3 & {\color{PineGreen} \small \bf ARM AZE GEO } \tabularnewline
		\hline
	%	{\color{RedOrange} \bf Community 6}& \bf 3 & {\color{RedOrange} \small \bf ISL MDA ROU} \tabularnewline
	%	 \hline
	\end{tabular}\caption{Members of the five most populous communities in $2019$.}
	\label{tab:CD2019BCG}
\end{table}

\begin{table}[h]
	\centering{}
	\begin{tabular}{|lll|}
		\hline \hline
		& \bf Size & \bf Members \tabularnewline
		{\color{red} \bf Community 1}  & \bf 6 & {\color{red} \small \bf AUS CAN JPN NZL PHL USA}\tabularnewline
		\hline
		{\color{blue} \bf Community 2} & \bf 12 & {\color{blue} \small \bf BEL CHE CZE DEU ESP GBR HUN} \tabularnewline
		{\color{blue} \bf } & \bf & {\color{blue} \small \bf IRL ITA NLD PRT SVK} \tabularnewline
		\hline                             
		{\color{orange} \bf Community 3} & \bf 5 & {\color{orange} \small \bf BIH HRV MNE SRB SVN}  \tabularnewline
		\hline
		{\color{ForestGreen} \bf Community 4}& \bf 7 & {\color{ForestGreen} \small \bf DNK EST FIN LTU LVA NOR SWE}	 \tabularnewline
		\hline
		{\color{PineGreen} \bf Community 5}& \bf 3 & {\color{PineGreen} \small \bf ARM AZE GEO } \tabularnewline
		\hline
		{\color{RedOrange} \bf Community 6}& \bf 3 & {\color{RedOrange} \small \bf ISL MDA ROU} \tabularnewline
		 \hline
	\end{tabular}\caption{Members of the six most populous communities in $2020$.}
	\label{tab:CD2020BCG}
\end{table}

Not surprisingly, we can observe a persistence of the community structure in WTN also during the pandemic situation. %It can be explained noticing three facts. The sample is composed by a low number of countries that, in addition, does not include China, which was the epicenter of the first wave of COVID-19. 
The type of data (average total trade exchanged) does not allow to show important movements. We expect that, focusing on some specific sectors (such as, for example, pharmaceutical industry), possible communities changes could be noticed but, at this time, the available data does not allow us to do it.\\ 
%Moreover, persistence in the WTN can be explained that 
Moreover, the analysis period starts from January and ends to June. Those months in $2020$ contain only the beginning of COVID-19 pandemic, i.e. the so called ``first wave''. %Focusing on the second reason, we notice that, 
Reasonably, the COVID-19 pandemic situation cannot be reflected immediately on the trade volumes as well as it is not possible to see the effects of the containment measures. 
%we expect that both factors affect the WTN from the second semester of $2020$ on. 

\begin{figure}[h]
\includegraphics[scale=0.05]{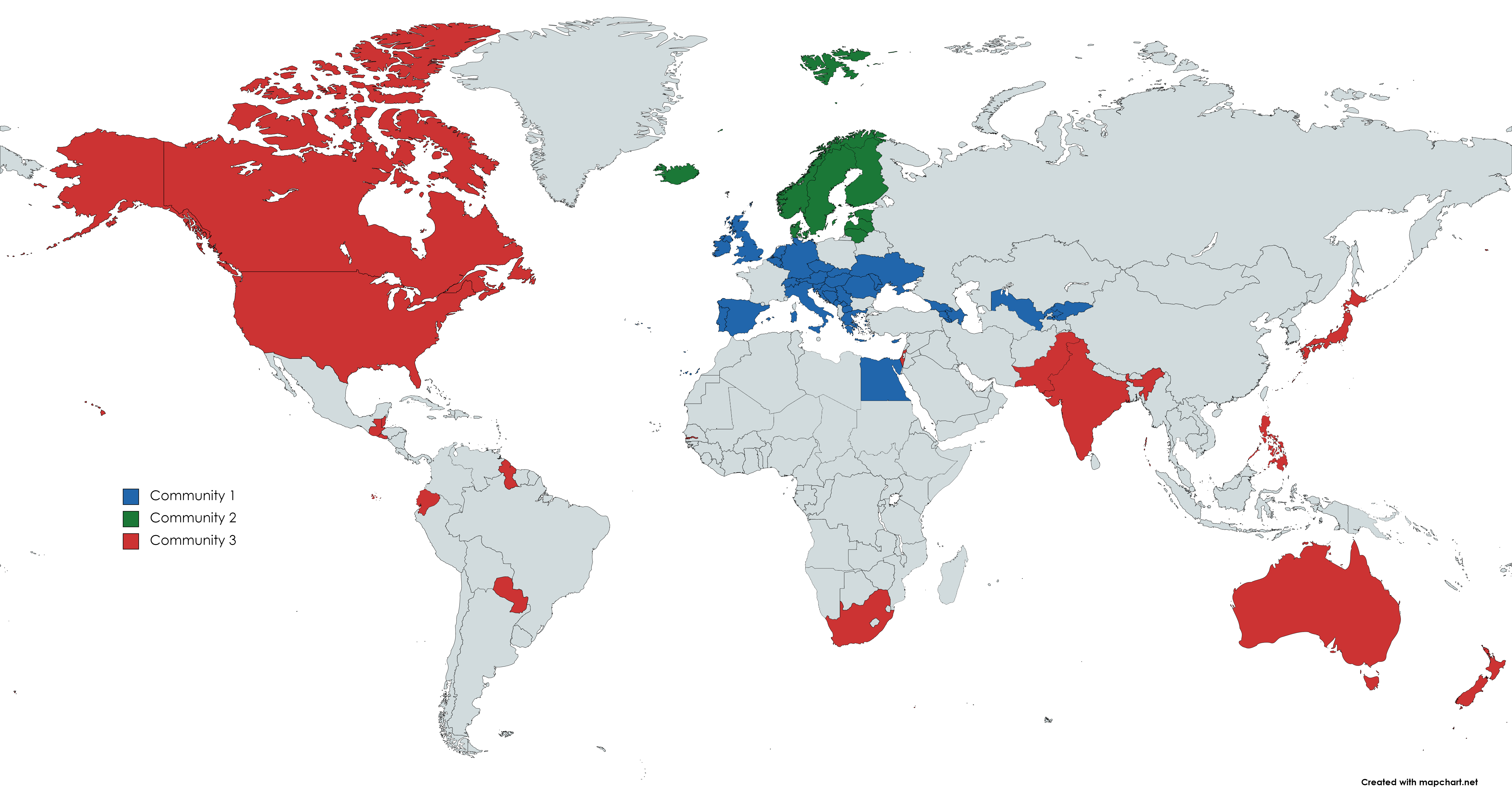}
\caption{World-map with the communities obtained with Louvain method (year $2019$).}
\label{figure_Louvain_2019}
\end{figure}

\begin{figure}[h]
\includegraphics[scale=0.05]{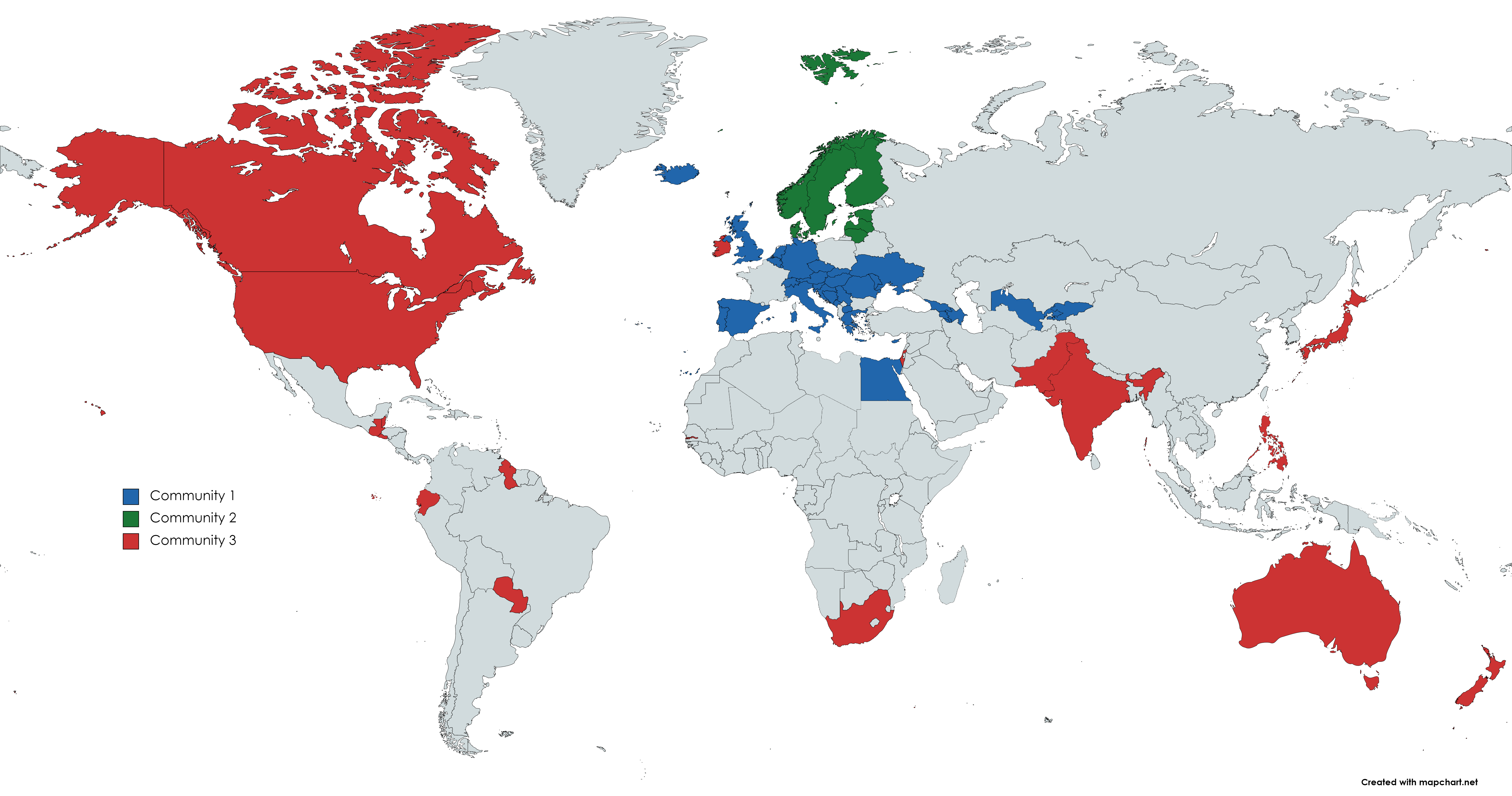}
\caption{World-map with the communities obtained with Louvain method (year $2020$).}
\label{figure_Louvain_2020}
\end{figure}

We now compare the results obtained with the methodology proposed in \cite{Bartesaghi2020} with a classical methodology in community detection, the Louvain method \cite{Blondel2008}. The method is based on the maximization of  a modularity score for each community, where the modularity function quantifies the quality of an assignment of nodes to communities. 
%that is the classical maximization of modularity using the Louvain method. For more explanations on the Louvain method we refer to \cite{Blondel2008}.\\
We observe that the classical method provides a less detailed division in the World Trade Network. In both year, we can observe three communities: the first one containing the Europe, the second one contains the USA and Pacific area and the third corresponds to the Northern Europe. Members of communities are plotted in Figures \ref{figure_Louvain_2019} and \ref{figure_Louvain_2020} for the years 2019 and 2020, respectively.\\

Members of communities obtained adopting the Louvain method are listed in Tables \ref{tab:Louvain_2019} and \ref{tab:Louvain_2020} for the years 2019 and 2020, respectively.  
%\color{fulvio} XXXCOMMENTO anche in questo caso, per riparmiare spazio sposterei le tabelle in appendiceXXX \color{black} \\

\begin{table}[h]
	\centering{}
	\begin{tabular}{|lll|}
		\hline \hline
		& \bf Size & \bf Members \tabularnewline
		{\color{blue} \bf Community 1}  & \bf 30 & {\color{blue} \small \bf ARM
AZE
BEL
BIH
CHE
CYP
CZE
DEU
EGY
ESP} \\
& & {\color{blue} \small \bf
GBR
GEO
GRC
HRV
HUN
IRL
ITA
KGZ
LUX
MDA} \\
& & {\color{blue} \small \bf
MKD
MNE
NLD
PRT
ROU
SRB
SVK
SVN
UKR
UZB
}
\tabularnewline
		\hline
		{\color{ForestGreen} \bf Community 2}& \bf 7 & {\color{ForestGreen} \small \bf DNK EST  ISL LTU LVA NOR SWE}	 \tabularnewline
		\hline
		{\color{Red} \bf Community 3}& \bf 18 & {\color{Red} \small \bf 
		 AUS
BLZ
CAN
ECU
GMB
GTM
GUY
IND
ISR
JPN} \\
& & {\color{Red} \small \bf 
NZL
PAK
PHL
PRY
SLV
USA
ZAF
} \tabularnewline
\hline
	\end{tabular}\caption{Members of the communities detected by Louvain method.}
	\label{tab:Louvain_2019}
\end{table}

\begin{table}[h]
	\centering{}
	\begin{tabular}{|lll|}
		\hline \hline
		& \bf Size & \bf Members \tabularnewline
		{\color{blue} \bf Community 1}  & \bf 30 & {\color{blue} \small \bf ARM
AZE
BEL
BIH
CHE
CYP
CZE
DEU
EGY
ESP}\\
& & {\color{blue} \small \bf
GBR
GEO
GRC
HRV
HUN
ISL
ITA
KGZ
LUX
MDA}\\
& & {\color{blue} \small \bf
MKD
MNE
NLD
PRT
ROU
SRB
SVK
SVN
UKR
UZB
}
\tabularnewline
		\hline
		{\color{ForestGreen} \bf Community 2}& \bf 7 & {\color{ForestGreen} \small \bf DNK EST FIN LTU LVA NOR SWE}	 \tabularnewline
		\hline
		{\color{Red} \bf Community 3}& \bf 18 & {\color{Red} \small \bf 
		 AUS
BLZ
CAN
ECU
GMB
GTM
GUY
IND
IRL
ISR} \\
& & {\color{Red} \small \bf 
JPN
NZL
PAK
PHL
PRY
SLV
USA
ZAF
} \tabularnewline
\hline
	\end{tabular}\caption{Members of the communities detected by Louvain method in $2020$.}
	\label{tab:Louvain_2020}
\end{table}

\begin{figure}[h]
\includegraphics[scale=0.4]{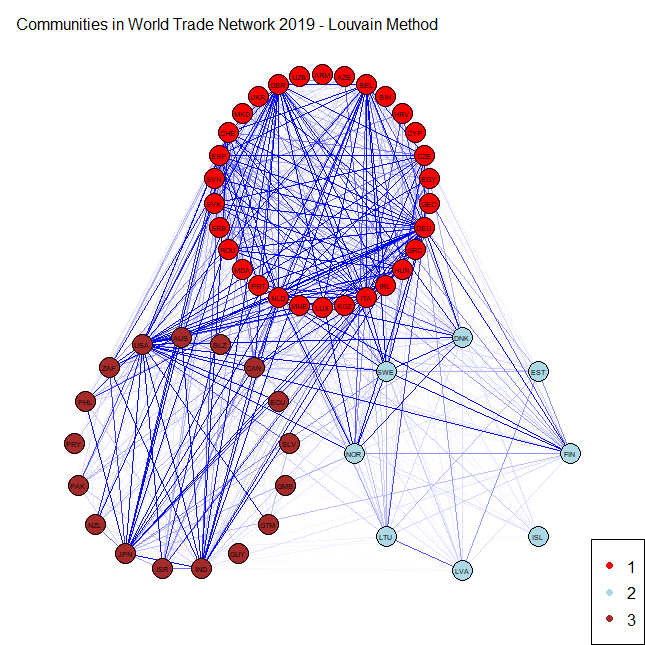}
\caption{Communities obtained with Louvain method referred to year $2019$.}
\label{m_figure_Louvain_2019}
\end{figure}

\begin{figure}[h]
\includegraphics[scale=0.4]{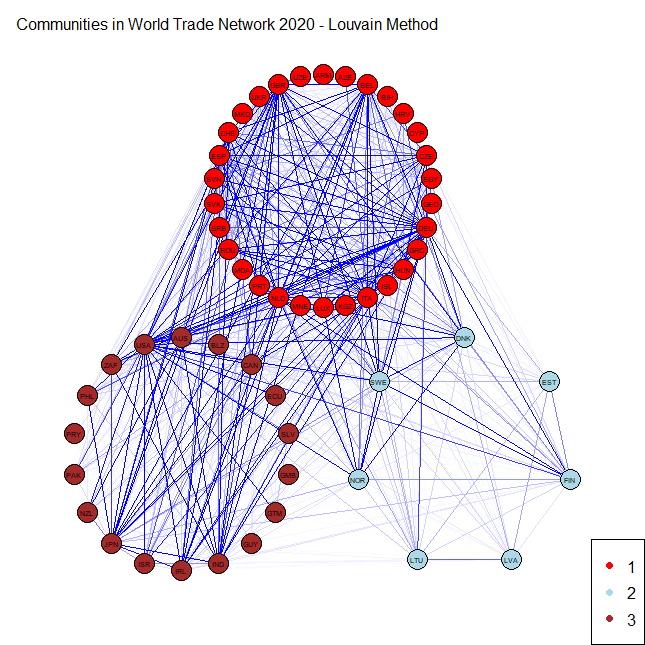}
\caption{Communities obtained with Louvain method referred to year $2019$.}
\label{m_figure_Louvain_2020}
\end{figure}

Figures \ref{m_figure_Louvain_2019} and \ref{m_figure_Louvain_2020} report the communities obtained applied the Louvain method for the year 2019 and 2020, respectively. \\

We notice that, as for the methodology based on communicability distance, also the Louvain method reveals a structural persistence in the WTN during the first wave of COVID-19 pandemic. This is in line with the recent results obtained by \cite{kiyota2021}.
We emphasize that the classical method catches only a persistence in the macroscopic structure while the method of \cite{Bartesaghi2020} reveals a persistence on the mesoscale structure. This result confirms that looking beyond the direct connections, it is possible to capture a strong interactions between countries.

\subsection{Impact of country's centrality measures on COVID-19 pandemic}

Tables \ref{NBR:inf} and \ref{NBR:deaths} show the results of regressions (as in Equation \eqref{econometric1}) concerning, respectively, the number of COVID-19 infections and the number of COVID-19 deaths. In both tables, each column reports the results of a regression based on a model that uses one-at-a-time the five different TNC indicators introduced in section \ref{econometric_model}.\\
From Table \ref{NBR:inf} we see that the IRRs of all our centrality indicators are always statistically significant and higher than 1: in general, a higher country centrality in the international trade network corresponds to a higher risk of infection. \\
Specifically, we find that, ceteris paribus, a one unit increase in each TNC indicator is associated to an expected increase in the risk of infection by a factor ranging from $3.2$ (Column 4 referred to weighted eigenvector) to $4.2$ (Column 3 refereed to local clustering coefficient). Moreover, we note that the IRR of each TNC indicator is always higher than the IRR of each other regressor, meaning that country's centrality in the world trade network is a key variable when analyzing the diffusion of COVID-19. \\
Interestingly, the AIC and BIC statistics, in line with the value of the pseudo log-likelihood and the pseudo $R^2$, show that the model with the highest goodness of fit with respect to the observed number of infections is that of Column 3, where the TNC indicator corresponds to the local clustering coefficient. Incidentally, the IRR of this latter is also the highest among all the other TNC indicators.\\

Looking at the other regressors, we find that the risk of infection increases with the country’s GDP per capita (Columns 3-5), with the share of elderly population (Column 1, and 3-5), and with a lower endowment of health facilities (Columns 3-5), confirming previous results obtained for a wider set of countries \cite{Antonietti2021}.\\

\begin{table}[h]
\centering
\begin{tabular}{l c c c c c} 
 \hline
DEP. VAR.: INF	&(1) &	(2)	&(3) &(4)	&(5) \\
\hline
Degree &	3.608*** & & & & \\				
      & (0.631)	& & & &\\			
Betweenness	& &	4.121***& & &\\			
         	& &	(1.868)	& & &\\		
Local Clustering   	& & &   	4.201***& &	\\	
            & & &	   	(0.968)	& &	\\
Weighted Eigenvector & &	& &			    3.225***& \\
			& & & &             (0.875)&	 \\
Strength (:$10^9$) & & & & &    			3.339*** \\
		        & & & & &    			(1.052) \\
GDPPC &	1.676&	1.147&	1.191**	&1.238*	&1.264**\\
 &    	(0.632)	&(0.266)&	(0.103)& (0.144)&	(0.133)\\
POP	&3.026	&1.813	&1.023	&1.132	&1.136\\
&	(3.227)&	(2.130)&	(0.134)&	(0.224)&	(0.184)\\
POP65+ 	&2.226***	&1.685	&1.688***&	2.680***&	2.420***\\
&	(0.685)	&(0.539)&	(0.294)	&(0.636)	&(0.564)\\
HBEDS&	0.678&	0.710&	0.585***&	0.557**	&0.555***\\
&	(0.212)&	(0.230)&	(0.087)	&(0.132)&	(0.107)\\
TEMP &	1.120&	0.728&	1.080&	1.387&	1.316\\
&	(0.226)&	(0.335)	&(0.142)&	(0.350)	&(0.231)\\
Week dummies & yes &yes &yes &  yes & yes\\
\hline
Overdispersion $(\alpha)$&	1.319***&	1.393***&	0.998***&	1.263***&	1.193***\\
&	(0.209)	&(0.170)&	(0.178)&	(0.213)&	(0.204)\\
\hline
N&	275&	275&	275&	275&	275\\
Log Pseudo-likelihood&	-2490.6	&-2500.5&	-2440.5&	-2482.5&	-2472.2\\
AIC&	5005.1&	5025.1	&4905.1	&4989.0	&4968.4\\
BIC	&5048.5	&5068.5	&4948.4	&5032.4	&5011.82\\
Pseudo $R^2$ &	0.076&	0.072&	0.094&	0.079&	0.083\\
Max VIF&	3.01&	3.20&	2.72&	2.57&	2.60\\
Mean VIF&	1.79&	1.89&	1.73&	1.69&	1.68\\
\hline
\hline

\multicolumn{6}{l}{\emph{Country-level clustered standard errors in parentheses.}}\\
\multicolumn{6}{l}{\emph{All the estimates also include a constant term. * $p < 0.1$, ** $p < 0.05$, *** $p < 0.01$.}}\\

\end{tabular}
\caption{Negative binomial regressions: infections, incidence rate ratios}
\label{NBR:inf}
\end{table}

%\color{fulvio} XXXDOVE? in geneale, bisognerebbe cercare di legare un po' il testo qui, anche solo con i riferimenti alle equazioni di ciascuna misura. una cosa del tipo, ... namely, Degree (Equation ???), Betweenness (Equation yyy), ecc. attenzione che bisogna mettre i numeri alle equazioni, che mancanoXXX  \color{black} \color{giorgio} concordo: non si capisce di quali equazioni si parla. Anche se credo che la scrittura delle variabili esplicative (simbolo e numero di equazione) debba essere fatto nella sezione del modello teorico quando si spiega la matrice TNC %TNC non é una matrice, ma un simbolo che rappresenta via via un diverso indicatore di centralità. L'alternativa, che non mi piace, é usare una equazione per ogni indicatore. Quindi TNC sta per indicatore di centralità, che via via sarà Degree, Eigenvector...

The same kind of results emerge for the case of COVID-19 fatalities, as shown in Table \ref{NBR:deaths}. Again, we find that the higher a country’s centrality in the trade network, the higher the risk of death due to COVID-19. Ceteris paribus, if a country's centrality increases by one unit, the risk of death is expected to increase by a factor ranging from $ 2.9$ (Column 1) to $8.6$ (Column 3). Again, the AIC and BIC statistics show that the econometric model that uses the local clustering coefficient is the one with the highest goodness of fit with respect to the observed number of deaths. \\
In addition, and still in line with previous literature \cite{Antonietti2021}, we find that the risk of death increases with the share of elderly population and with a lower endowment of hospital beds in a country. On top of this, both in Table \ref{NBR:inf} and in Table \ref{NBR:deaths}, the VIF statistics are low enough with respect to the commonly accepted threshold of 5, showing again that multicollinearity is not an issue. \\

\begin{table}[h]
\centering
\begin{tabular}{l c c c c c} 
 \hline
DEP. VAR.: DEATH	&(1) &	(2)	&(3) &(4)	&(5) \\
\hline
Degree &	2.878***	& & & & \\			
&	(0.888)	& & & & \\			
Betweenness	&&	6.750***	& & & \\		
&&		(3.022)	& & & \\		
Local Clustering	&&&		8.648***	& &	\\		
		&&&	(3.579)		& &	\\	
Weighted Eigenvector &&&&				5.123**	& 	\\	
			&&&&	(3.339)	&	\\	
Strength (:$10^9$)	&&&&&				6.893***\\
				&&&&&	(5.123)\\
GDPPC&	1.458&	0.755&	0.930&	0.933&	0.966\\
&	(0.671)	&(0.162)&	(0.141)	&(0.149)&	(0.136)\\
POP&	2.897	&1.486	&0.867&	1.020	&1.016\\
&	(3.258)&	(1.636)&	(0.102)	&(0.196)&	(0.169)\\
POP65+	&6.223***	&3.099***&	2.136***&	5.423***&	4.037***\\
&	(2.549)	&(1.216)&	(0.606)&	(2.338)&	(1.830)\\
HBEDS&	0.370***&	0.434***&	0.484***&	0.382***&	0.429***\\
&	(0.105)	&(0.127)&	(0.099)	&(0.119)&	(0.127)\\
TEMP&	1.721&	0.973&	1.426*&	1.968&	1.915**\\
&	(0.471)	&(0.424)&	(0.272)&	(0.814)&	(0.556)\\
Week dummies & yes &yes&yes&yes& yes\\
\hline
Overdispersion ($\alpha$)&	2.245***&	2.055***&	1.400***&	1.933***&	1.805***\\
&	(0.283)&	(0.255)&	(0.217)&	(0.266)&	(0.263)\\
\hline
N&	275	&275&	275&	275&	275\\
Log Pseudo-likelihood&	-1519.4	&-1500.6&	-1437.9	&- 1490.9&	-1479.2\\
AIC&	3062.9&	3025.3&	2899.9	&3005.7	&2982.3\\
BIC&	3106.3&	3068.7&	2943.3&	3049.2&	3025.7\\
Pseudo R2&	0.101&	0.112&	0.149&	0.118&	0.124\\
Max VIF	&3.01&	3.20&	2.72&	2.57&	2.60\\
Mean VIF&	1.79&	1.89&	1.73&	1.69&	1.68\\
 \hline
\hline
\multicolumn{6}{l}{\emph{Country-level clustered standard errors in parentheses.}}\\
\multicolumn{6}{l}{\emph{All the estimates also include a constant term. * $p < 0.1$, ** $p < 0.05$, *** $p < 0.01$.}}\\

\end{tabular}
\caption{Negative binomial regressions: deaths, incidence rate ratios}
\label{NBR:deaths}
\end{table}

As an additional step in the analysis, we re-estimate Equation \eqref{econometric1} using a community-specific (based on \cite{Bartesaghi2020}) averaged local clustering coefficient and comparing the IRRs with those estimated for the local clustering coefficient (as in Equation \eqref{onnela_meas}) shown in column $3$ of Tables \ref{NBR:inf} and \ref{NBR:deaths}. \\
Specifically, each country belongs to one of the $22$ communities identified by the methodology in \cite{Bartesaghi2020}. For each community, we compute the average of the clustering coefficients $\bar{C}$ of the countries therein. Therefore, each country in community $k$ has a new clustering coefficient equal to $\bar{C}_{k}$, defined by
\begin{equation}
\bar{C}_{k} = \frac{1}{n_k} \sum_{j=1}^{n_k} C_{j}
\label{onnela_medio}
\end{equation}
where $n_k$ is the size of community $k$ and $C_{i}$ is the local clustering coefficient of node $i$ as in Equation \eqref{onnela_meas}.\\
Coefficient $\bar{C}$ has two properties: on the one hand, it still reflects the country's centrality within 
all its triadic relations expressed by the local clustering coefficient in Equation \eqref{onnela_meas}. On the other hand, $\bar{C}$ takes into account the mesoscale structure of the WTN based on communicability. In other words, with this new coefficient, we capture the impact of a country centrality in a subset of the world trade network, where nodes strongly exchange trade-related information that can be directly observable (such as merchandise trade) or indirectly observable (such as the interactions characterizing the supply chain of a good). \\

Then, we re-estimate Equation \eqref{econometric1} using as network centrality measure the average community local clustering $\bar{C}_k$ as in Equation \eqref{onnela_medio}, and we compare the newly estimated IRR with that of the local clustering coefficient in Tables \ref{NBR:inf} and \ref{NBR:deaths}. The results are shown in Table \ref{confrontoOnnela}. 

Columns 1 and 3 report the results shown in Column 3 of Tables \ref{NBR:inf} and \ref{NBR:deaths} concerning infections (INF) and deaths (DEATH), respectively, while Columns 2 and 4 show the new results for the model that uses average community coefficient $\bar{C}_k$ as main regressor.
%{\bf old}
%While Columns 2 and 4 show the new results when the local clustering indicator is computed as a community-specific average.

Interestingly, we find that a unit increase in averaged local clustering coefficient $\bar{C}_k$ corresponds to a higher risk of infection and death as compared to country $i$ local clustering coefficient $C_i$. Those risks pass from an order of $4.2$ to $5.6$ in the case of infection and from an order of $8.6$ to $10.3$ in the case of death. These results confirm that community-specific measures of country centrality can provide even stronger results on the diffusion patterns of COVID-19. The fact that communities are detected using a wider set of trade-related information between countries, which implicitly include unobserved flows of people other than merchandise, allows accounting for a higher risk of contagion attributable to international trade.

\begin{table}[h]
\centering
\begin{tabular}{l c c |c c} 
 \hline
DEP. VAR.: & \multicolumn{2}{c|}{INF} & \multicolumn{2}{c}{DEATH}\\
\hline \\
& (1) & (2) & (3) &(4)\\
\hline
Local Clustering &	4.201*** &	& 8.648***	& \\
       & 	(0.968)	& & 	(3.579)	& \\
Average Community Clustering	& &	5.551*** & &	10.25***\\
 & &		(2.486)	& & 	(4.977)b\\
GDPPC &	1.191**	& 1.510** &	0.930 &	1.166 \\
 &  	(0.103) &	(0.243) &	(0.141)	&(0.199)\\
POP &	1.023 &	2.499     &	0.867   & 	2.107 \\
 &  	(0.134)&	(2.467)&	(0.102)&	(2.137)\\
POP65+ &1.688***&	1.705**	&2.136*** &	2.945*** \\
&	(0.294) &	(0.425)	&(0.606) &	(0.904)\\
HBEDS &	0.585*** &	0.652* &	0.484***&	0.419***\\
 & 	(0.087) &	(0.146)	 &(0.099) &	(0.087)\\
TEMP &	1.080 &	0.980 &	1.426* &	1.294\\
 & 	(0.142) &	(0.286)	& (0.272) &	(0.448)\\
Week dummies & yes & yes & yes & yes\\
\hline
Overdispersion ($\alpha$) &	0.998*** &	1.344*** &	1.400***&	1.898***\\
 &	(0.178)	&(0.211) &	(0.217) &	(0.284) \\
\hline
N &	275 &	275	& 275 &	275\\
Pseudo $R^2$	&0.094	&0.074	&0.149	&0.120\\
\hline
\hline
\multicolumn{5}{l}{\emph{Country-level clustered standard errors in parentheses.}}\\
\multicolumn{5}{l}{\emph{All the estimates also include a constant term.}}\\
\multicolumn{5}{l}{\emph{* $p < 0.1$, ** $p < 0.05$, *** $p < 0.01$.}}\\

\end{tabular}
\caption{General vs average community clustering coefficient and COVID-19 diffusion.}
\label{confrontoOnnela}
\end{table}

\section{Conclusions}

In this paper, we evaluate the relationship between the WTN structure and the COVID-19 disease. 
The complex structure of the trade relationships between countries requires to be investigated using effective network tools, to reveal its hidden mesoscale structure, characterised by strong interconnections, as well as to assess countries' central position in the network.

%The Trade network has a complex structure, that can be studied showing how trade relationships across countries give rise to homogeneous cluster of countries, characterised by strong trade connections, as well as on focusing on countries' centrality in the WTN structure. 
Those trade relationships could have been impacted by the COVID-19 pandemic, both directly because of the spread-out of the virus and indirectly due to the policies that countries have implemented in order to reduce the pandemic diffusion and consequences. At the same time, it is well possible that the pandemic itself has been favoured by the complex network of relationships that take place when trade occurs.
Through network measures we have evaluated %both globally and at the level of countries' clustering, 
to what extend the WTN has been affected by the COVID-19, and to what extent countries' centrality explains the diffusion and mortality of COVID-19. 
Moreover, we have shown that the WTN mesoscale structure has been resilient to the diffusion of the pandemic. Even if such a result was expected at the global level, we have shown that it holds also when looking at the number and members of the communities that emerged before and during the outbreak of the COVID-19, showing that the strength of long-range alliances have not been affected by the beginning of the COVID-19 pandemic. \\
On the contrary, the country centrality has shown to be a key explanatory variable for the diffusion and mortality of COVID-19. We showed that country centrality measures strongly explained the risk of infection and mortality, when controlling for other possible confounding socio-economic factors.\\
Both results can be of interest for the analysis of the structure and evolution of the WTN and from the point of view of studying the determinants and the consequences of the COVID-19 pandemic. Moreover, establishing the link between a pandemic and the structure of the network of international trade can provide useful policy insights. More precisely, this knowledge can guide decision makers about the adoption and calibration of relevant public safety policies, such as general lockdown measures and temporary trade bans, which have huge economic consequences but have shown so far unclear effectiveness on the virus spread-out. This can be important for both the actual COVID-19 pandemic, which at the time of the writing of this article is still widely diffused worldwide, as well as for the unfortunate yet possible case of the diffusion of a future new pandemic event.  

\bibliography{References}

\appendix
\section{Table of countries}
\begin{table}[h]
    \centering
    \begin{tabular}{ll}
          Contry &	Code\\
          \hline
Armenia	&	ARM	\\
Australia	&	AUS	\\
Azerbaijan	&	AZE	\\
Belgium	&	BEL	\\
Belize	&	BLZ	\\
Bosnia Herzegovina	&	BIH	\\
Canada	&	CAN	\\
Croatia	&	HRV	\\
Cyprus	&	CYP	\\
Czech Rep.	&	CZE	\\
Denmark	&	DNK	\\
Ecuador	&	ECU	\\
Egypt	&	EGY	\\
El Salvador	&	SLV	\\
Estonia	&	EST	\\
Finland	&	FIN	\\
Gambia	&	GMB	\\
Georgia	&	GEO	\\
Germany	&	DEU	\\
Greece	&	GRC	\\
Guatemala	&	GTM	\\
Guyana	&	GUY	\\
Hungary	&	HUN	\\
Iceland	&	ISL	\\
India	&	IND	\\
Ireland	&	IRL	\\
Israel	&	ISR	\\
    \end{tabular}
    \caption{Table of countries I }
    \label{tab:countries1}
\end{table}

\begin{table}[h]
    \centering
    \begin{tabular}{ll}
          Contry &	Code\\
          \hline
Italy	&	ITA	\\
Japan	&	JPN	\\
Kyrgyzstan	&	KGZ	\\
Latvia	&	LVA	\\
Lithuania	&	LTU	\\
Luxembourg	&	LUX	\\
Montenegro	&	MNE	\\
Netherlands	&	NLD	\\
New Zealand	&	NZL	\\
Norway	&	NOR	\\
Pakistan	&	PAK	\\
Paraguay	&	PRY	\\
Philippines	&	PHL	\\
Portugal	&	PRT	\\
Rep. Of Moldova	&	MDA	\\
Romania	&	ROU	\\
Serbia	&	SRB	\\
Slovakia	&	SVK	\\
Slovenia	&	SVN	\\
South Africa	&	ZAF	\\
Spain	&	ESP	\\
Sweden	&	SWE	\\
Switzerland	&	CHE	\\
TFYR of Macedonia	&	MKD	\\
Ukraine	&	UKR	\\
United Kingdom	&	GBR	\\
United States of America	&	USA	\\
Uzbekistan	&	UZB	\\

    \end{tabular}
    \caption{Table of countries II.}
    \label{tab:countries2}
\end{table}

\end{document}